\documentclass[sigconf]{acmart}
\AtBeginDocument{%
  \providecommand\BibTeX{{%
    \normalfont B\kern-0.5em{\scshape i\kern-0.25em b}\kern-0.8em\TeX}}}

\copyrightyear{2023}
\acmYear{2023}
\setcopyright{rightsretained}
\acmConference[FAccT '23]{2023 ACM Conference on Fairness, Accountability, and Transparency}{June 12--15, 2023}{Chicago, IL, USA}
\acmBooktitle{2023 ACM Conference on Fairness, Accountability, and Transparency (FAccT '23), June 12--15, 2023, Chicago, IL, USA}
\acmDOI{10.1145/3593013.3593978}
\acmISBN{979-8-4007-0192-4/23/06}

\acmSubmissionID{43}

\usepackage{gensymb}
\usepackage{booktabs}
\usepackage{xcolor,colortbl}
\usepackage{tabu}
\usepackage{arydshln}
\usepackage{multicol, multirow}
\usepackage{doi}
\usepackage{url}

\usepackage[capitalize]{cleveref}
\crefname{section}{Sec.}{Secs.}
\Crefname{section}{Section}{Sections}
\Crefname{table}{Table}{Tables}
\crefname{table}{Tab.}{Tabs.}

\newcommand{\interviewprefix}{P}
\makeatletter
\newcommand{\interview}[1]{%
  \def\nextitem{\def\nextitem{, }}%
  (\@for\el:=#1\do{\nextitem\interviewprefix\el})%
}
\newcommand{\shortquote}[1]{“\textit{#1}”}%

\begin{document}

\title[Understanding End-Users’ Trust in a Real-World Computer Vision Application]{Humans, AI, and Context: Understanding End-Users’ Trust in a Real-World Computer Vision Application}

\author{Sunnie S. Y. Kim}
\orcid{}
\affiliation{%
  \institution{Princeton University}
  \city{Princeton}
  \state{New Jersey}
  \country{USA}
}

\author{Elizabeth Anne Watkins}
\authornote{Most work was done during Postdoctoral Research Appointment at the Princeton Center for Information Technology Policy and Human-Computer Interaction Group.}
\affiliation{
  \institution{Intel Labs}
  \city{Santa Clara}
  \state{California}
  \country{USA}
}

\author{Olga Russakovsky}
\affiliation{%
  \institution{Princeton University}
  \city{Princeton}
  \state{New Jersey}
  \country{USA}
}

\author{Ruth Fong}
\affiliation{%
  \institution{Princeton University}
  \city{Princeton}
  \state{New Jersey}
  \country{USA}
}

\author{Andrés Monroy-Hernández}
\affiliation{%
  \institution{Princeton University}
  \city{Princeton}
  \state{New Jersey}
  \country{USA}
}

\renewcommand{\shortauthors}{Kim, Watkins, Russakovsky, Fong, Monroy-Hernández}

\begin{abstract}
Trust is an important factor in people's interactions with AI systems. However, there is a lack of empirical studies examining how real end-users trust or distrust the AI system they interact with. Most research investigates one aspect of trust in lab settings with hypothetical end-users. In this paper, we provide a holistic and nuanced understanding of trust in AI through a qualitative case study of a real-world computer vision application. We report findings from interviews with 20 end-users of a popular, AI-based bird identification app where we inquired about their trust in the app from many angles. We find participants perceived the app as trustworthy and trusted it, but selectively accepted app outputs after engaging in verification behaviors, and decided against app adoption in certain high-stakes scenarios. We also find domain knowledge and context are important factors for trust-related assessment and decision-making. We discuss the implications of our findings and provide recommendations for future research on trust in AI.
\end{abstract}

\ccsdesc[500]{Human-centered computing~Empirical studies in HCI}
\ccsdesc[500]{Human-centered computing~User studies}
\ccsdesc[500]{Computing methodologies~Artificial intelligence}

\keywords{Trust in AI, Human-AI Interaction, Computer Vision, Case Study}

\maketitle

\section{Introduction}
\label{sec:intro}

Trust is an important factor in people's interactions with Artificial Intelligence (AI) systems.
For the effective adoption and use of these systems, people must trust them appropriately.
Both unwarranted trust (trusting when the AI system is not trustworthy) and unwarranted distrust (distrusting when the AI system is trustworthy) can hurt the quality of interactions~\cite{wischnewski2023chi,Jacovi2021FAccT,Miller2022TRAIT,Banovic2023CSCW}.
To better understand trust and foster it appropriately in human-AI interactions, recent works have started to investigate questions such as: 
What does it mean to trust an AI system?~\cite{Jacovi2021FAccT,Toreini2020FAccT,Glikson2020review}
How is trust established and developed?~\cite{Liao2022MATCH,Kahr2023IUI}
What factors influence people's trust and how?~\cite{Zhang2020FAccT,Yin2019trust,Lai2019trust,Cheng2019CHI,Schaffer2019trust,Yu2019IUI,Hartmann2022AMCIS}.

Trust in AI research, however, is still in a nascent stage.
As noted in recent surveys~\cite{Ueno2022CHIEA,Vereschak2021CSCW},
papers often use different definitions of trust, making their results difficult to compare.
There is also little agreement on how to empirically study trust, e.g., when to use subjective vs. objective measures.
Finally, there is a lack of research that approaches trust holistically.
Most papers study one specific aspect of trust (e.g., whether explainability increases people's trust in AI~\cite{Schaffer2019trust,Cheng2019CHI,Kocielnik2019CHI,Lai2019trust,zhang2016EB,Bansal2021Team,Kim2022HIVE,nguyen2021neurips,Nguyen2022team,Yin2019trust,Bucinca2020IUI}) in lab settings with hypothetical end-users.
While they provide valuable insights, they do not capture the complex nuances of trust in real-world contexts.

The FAccT community is increasingly focusing on trust because understanding and measuring it overlap with evaluating fairness, accountability, and transparency of algorithmic systems.
Work around those principal interests has begun to mature, as scholars begin to conduct more empirical research on factors and human perceptions of fairness \cite{dodge2019explaining,harrison2020empirical,kasinidou2021agree}, and going even further, to recognize the influence of sociotechnical context on factors of fairness, ethics, and accountability~\cite{selbst2019fairness,green2020algorithmic,metcalf2021algorithmic,sambasivan2021re,sloane2022german}. However, papers on trust recently published at FAccT remain largely theory-focused~\cite{Liao2022MATCH,Ferrario2022FAccT,Thornton2022FAccT,Toreini2020FAccT,Jacovi2021FAccT,Knowles2021FAccT}. Our work contributes to the community by maturing the understanding of trust through an in-depth empirical study, moving the conversation as AI systems move from theoretical, lab-based projects out into the real world.

\begin{figure*}[t!]
\centering
\includegraphics[width=0.59\linewidth]{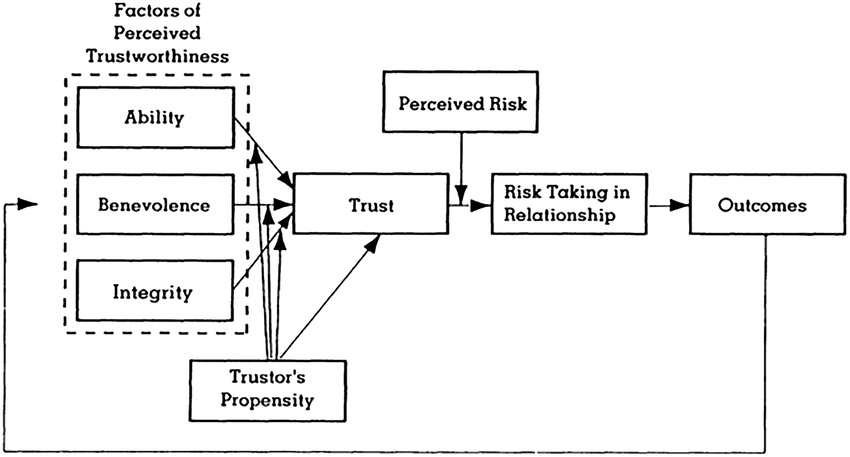}
\caption{Based on Mayer et al.'s trust model \cite{Mayer1995trust}, we separate \textit{trust} from \textit{trustworthiness} perceptions that precede it, and two trust-related behaviors that proceed from it: 
(1) AI \textit{adoption}, which corresponds to \textit{risk taking in relationship} in the model, and (2) AI \textit{output acceptance}, which corresponds to \textit{outcomes} evaluation in the model.
In this work, we describe both \textit{general} trustworthiness perceptions and trust attitudes, and \textit{instance-specific} trust-related behaviors.
See \cref{sec:rw_definitions} for further discussion. This figure is replicated from Mayer et al.'s work \cite{Mayer1995trust}.
}
\label{fig:mayer}
\end{figure*}

The goal of this work is to provide a more holistic and nuanced understanding of trust in AI through a qualitative case study of a real-world AI application.
We ground our study in Merlin~\cite{merlin}, a free mobile phone app that uses computer vision AI models to identify birds in user-uploaded photos and audio recordings (\cref{fig:merlin_ui}). 
We chose Merlin because it is a widely-used app that allows us to connect with a diverse set of active end-users with varying levels of domain (bird) and AI background, and satisfies the requirements of commonly-used trust definitions~\cite{Mayer1995trust,LeeAndSee2004}.
Concretely, we conducted semi-structured interviews with 20 Merlin end-users and inquired about their trust in the app from several angles.
\footnote{In the same interviews, we also inquired about participants' AI explainability needs, intended uses of AI explanations, and perceptions of existing explanation approaches, and analyzed that portion of the data in another paper~\cite{kim2023helpmehelptheai}.}
Since we were one of the first to talk to actual end-users about their trust relationship with the AI application, we focused on exploring what factors influence trust and how, rather than quantifying the importance of certain pre-specified factors.
Afterwards, we applied Mayer et al.'s theoretical definition and model of trust~\cite{Mayer1995trust} (\cref{fig:mayer}) to the collected empirical data, to delineate and describe multiple aspects of trust and their influencing factors.

We make three key contributions in this work:
(1) We further the FAccT community's understanding of trust, to date still heavy in theory, through an in-depth qualitative empirical study. Specifically, we study how end-users trust AI in a real-world context and what factors influence their trust. In doing so, we synthesize theoretical and empirical trust research by applying Mayer et al.'s theory~\cite{Mayer1995trust} to the empirical data we collected. This approach will, we hope, yield insights into how readily existing theories of trust can be operationalized for empirical research.
(2) We provide a more holistic and nuanced understanding of trust, as compared to the current state-of-the-art in the field.
We find general trustworthiness perceptions and trust attitudes are distinct from instance-specific trust-related behaviors.
While our participants told us they perceived the app as trustworthy and trusted it, they also described how they selectively accepted the app outputs after engaging in verification behaviors, and sometimes decided against app adoption in certain high-stakes scenarios. 
Domain knowledge and context were particularly important factors for participants' trust-related assessment and decision-making.
(3) Finally, we discuss the implications of our findings and provide practical recommendations for future research on trust in AI. Most critically, we advocate for researchers to define and delineate trust from related constructs, and to consider human, AI, and context-related factors of trust together.

\section{Background and related work}
\label{sec:relatedwork}

\subsection{Definitions and models of trust in AI}
\label{sec:rw_definitions}

Trust has a long history of research. Before ``trust in AI'' was researched, disciplines ranging from psychology to philosophy to human factors had studied trust in the context of relationships between humans or between humans and machines. As such, there are numerous definitions and models of trust, but most have their roots in two papers: Mayer et al.'s 1995 paper on organizational trust~\cite{Mayer1995trust} and Lee and See's 2004 paper on trust in automation~\cite{LeeAndSee2004}.

Within trust in AI research, many works do not state a definition of trust, according to recent review papers of the field~\cite{Vereschak2021CSCW,Ueno2022CHIEA,Glikson2020review}.
Among works that do, the most commonly used definitions come from the aforementioned papers: (1) ``the willingness of a party to be vulnerable to the actions of another party based on the expectation that the other will perform a particular action important to the trustor, irrespective of the ability to monitor or control that party'' by Mayer et al.~\cite{Mayer1995trust} and (2) ``the attitude that an agent will help achieve an individual’s goals in a situation characterized by uncertainty and vulnerability'' by Lee and See~\cite{LeeAndSee2004}.

Both definitions share the same key elements, as described by Vereschak et al.~\cite{Vereschak2021CSCW}:
(1) \textit{vulnerability}: the situation involves uncertainty of outcomes and potential negative consequences; 
(2) \textit{positive expectations}: the trustor thinks that negative outcomes associated with trusting do not exist or are very unlikely;
and (3) \textit{attitude}: the general way of thinking and feeling, typically reflected in a behavior, although not a behavior itself.
These elements also distinguish trust from other related constructs.
It is not \textit{trust} but:
\textit{confidence} when there is no vulnerability;
\textit{distrust} when there is no positive expectation;
\textit{compliance} or \textit{reliance} when referring to a behavior;
and \textit{perceived trustworthiness} when referring to a perception of a trustee's characteristics upon which trustors form their trust.
For example, when a study measures whether participants follow AI's advice, which is a directly observable behavior, it is measuring \textit{reliance} not \textit{trust}. When a study asks participants to rate their trust level on a survey scale, it is measuring attitude, except when there is no vulnerability (e.g., lab experiment with no incentives or risks), it is measuring \textit{confidence} not \textit{trust}. 
We hope these distinctions reduce ambiguity and confusion around the terms.

Recently, scholars have proposed specific definitions and models for ``trust in AI''~\cite{Jacovi2021FAccT,Liao2022MATCH}.
Jacovi et al.~\cite{Jacovi2021FAccT} formalized trust in AI as ``contractual'': to trust an AI system is to believe that it is trustworthy to uphold some contract. Their formalization disentangles trust and trustworthiness, and defines ``warranted trust'' as trust that is ``caused'' by the AI's trustworthiness. 
Liao and Sundar~\cite{Liao2022MATCH}, on the other hand, took a communication perspective and proposed a model that describes how the trustworthiness of AI systems is communicated through trustworthiness cues and how those cues are processed by people to make trust judgments.

In this work, we adopt the model of trust by Mayer et al.~\cite{Mayer1995trust}, despite it being developed for organizational trust, because the model's definition and process orientation fit our work's objective of holistically understanding trust, in the context of human-AI interaction.
Mayer et al. \cite{Mayer1995trust} delineate trust from its antecedents, context, and products, and describe how different components influence each other as the trustor interacts with the trustee (\cref{fig:mayer}).
Based on their model, we separate trust from trustworthiness perceptions (trustworthiness being trust's antecedent) and trust-related behaviors, i.e., output acceptance and adoption decisions. The models by Jacovi et al. and Liao and Sundar~\cite{Jacovi2021FAccT,Liao2022MATCH} are less fitting for our work because Jacovi et al. \cite{Jacovi2021FAccT} focus on formalizing prerequisites, causes, and goals of trust in AI, and Liao and Sundar \cite{Liao2022MATCH} focus on modeling the communication of trustworthiness.

\subsection{Empirical studies of trust in AI}
\label{sec:rw_empricial}

Trust in AI is a fast-growing research field with significant empirical work.
However, there is surprisingly little research on how end-users trust AI in \textit{real-world contexts}, and \textit{what factors} influence their trust---two gaps our work aims to fill.
Filling these gaps is an important endeavor because while AI systems may perform well in controlled lab settings, their take-up and use in the real world are subject to various factors, many of which are context-dependent and currently under-anticipated in research and system design.

Much of prior work focus on understanding the effect of certain \textit{pre-specified factors} on trust~\cite{Zhang2020FAccT,Yin2019trust,Lai2019trust,Cheng2019CHI,Schaffer2019trust,Yu2019IUI,poursabzi2021manipulating,nguyen2021neurips,Nguyen2022team,Kim2022HIVE,Bansal2021Team,Kocielnik2019CHI,Hartmann2022AMCIS,Wang2008ecommerce}.
Most utilize \textit{lab experiments}, usually with participants recruited from crowdsourcing platforms (e.g., MTurk~\cite{Zhang2020FAccT,Yin2019trust,Lai2019trust,Cheng2019CHI,Schaffer2019trust,poursabzi2021manipulating,nguyen2021neurips,Nguyen2022team,Kim2022HIVE,Bansal2021Team}, Prolific~\cite{nguyen2021neurips,Nguyen2022team}, internal platform~\cite{Kocielnik2019CHI}).
Typically, these works start with a hypothesis (e.g., explainability will increase trust in AI).
To investigate the hypothesis, they choose a measure of trust (e.g., self-reported rating on a 1-7 scale~\cite{Jian2000Scale}), make a change to the factor of interest in the design of the AI system (e.g., show an explanation of the AI's output), and then quantify the effect of that change on participants' trust.
Based on the results, they conclude the effect of the factor of interest on trust.

The most commonly studied factors in the literature are transparency and explainability.
However, researchers operationalize these factors in several ways.
For instance, \textit{transparency} is operationalized as providing model internals (e.g., learned coefficients in a linear regression model) in \cite{poursabzi2021manipulating}, overall performance measures (e.g., accuracy) in \cite{Schaffer2019trust,Lai2019trust,Yu2019IUI,Yin2019trust,Kocielnik2019CHI}, confidence scores on individual outputs in \cite{Zhang2020FAccT,nguyen2021neurips}, and visualizations of input data distributions and feature engineering process in \cite{Drozdal2020AutoML}.
Similarly, while \cite{Schaffer2019trust,Cheng2019CHI,Kocielnik2019CHI,Lai2019trust,zhang2016EB,Bansal2021Team,Kim2022HIVE,nguyen2021neurips,Nguyen2022team,Yin2019trust,Bucinca2020IUI} all study the effect of \textit{explainability} on trust, the operationalized explanations of AI's behavior and outputs greatly vary in approach (e.g., feature attribution, counterfactual examples) and form (e.g., heatmap-based, part-based).

These works provide insights into the relationship between trust and the factor of interest, as operationalized in a specific and controlled way. 
However, they do not capture the contextual aspects of trust, and the design of these studies does not allow for discovering new trust-influencing factors.
To address these two gaps, we conducted a qualitative case study of a real-world computer vision AI application and interviewed its end-users about their trust relationship with the AI.
While resource-intensive, interviews enabled us to explore multiple aspects of trust in depth and identify trust-influencing factors in a bottom-up manner.
The value of qualitative case studies has been demonstrated in recent works~\cite{Sendak2020SepsisWatch,elish2020repairing,Widder2021CHI}. In one example, Widder et al.~\cite{Widder2021CHI} conducted a case study investigating what factors influence engineers' trust in an autonomous software engineering tool in a high-stakes workspace.
They found that trust, in their study setting, was influenced by the tool's transparency, usability, social context, and the organization's associated processes. Widder et al.'s work laid groundwork for our own qualitative study, as we applied their methods to ask similar questions about a different population in a different domain: end-users of an AI-based bird identification app.
Our work provides complementary insights, and we encourage the community to conduct more qualitative case studies of trust in AI.

\section{Methods}
\label{sec:methods}

In this section, we describe our study methods. All were reviewed and approved by our Institutional Review Board.

\subsection{Study application: Merlin Bird ID app}
\label{sec:researchcontext}

\begin{figure*}[t!]
\centering
\includegraphics[width=\linewidth]{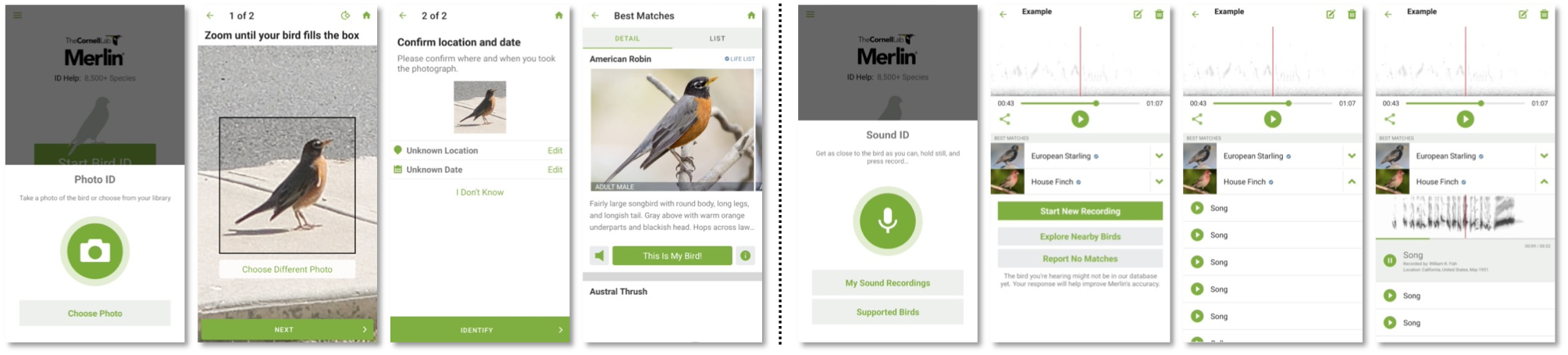}
\caption{Our study application Merlin~\cite{merlin} is a popular AI-based bird identification mobile phone app. Users upload photos on the Photo ID feature (left) or audio recordings on the Sound ID feature (right), with optional location and season data, and get a list of bird(s) that best match the input. See \cref{sec:researchcontext} for more details about the app.}
\label{fig:merlin_ui}
\end{figure*}

To study trust in AI in a realistic setting, we looked for a research setting that first, involves real-world AI use by end-users who range in their domain and AI background, and second, satisfies the requirements of widely-accepted trust definitions~\cite{Mayer1995trust,LeeAndSee2004}.
We found Merlin~\cite{merlin} (\cref{fig:merlin_ui}) to satisfy both conditions.
First, Merlin is a mobile phone app that identifies bird species from user-input photos and/or audio recordings. It is an \textit{expert} application with expertise that most people do not have, i.e., knowledge and skill to identify thousands of birds.  As a free app with over a million downloads, it is used by people with diverse domain (bird) and AI backgrounds, thus satisfying our first requirement. 
Second, while Merlin is generally a \textit{low-stakes} application, there are some amount of \textit{vulnerability} and \textit{positive expectations} in its use, as we verify in \cref{sec:prelude}. This allows us to characterize end-users' \textit{attitude} toward the app as trust, and study ``trust in AI'' and its influencing factors.

\subsection{Participant recruitment and selection}
\label{sec:recruitment}

We recruited participants who are active end-users of Merlin Photo ID or Sound ID, the app's AI-based bird identification features, with considerations for diversity in the domain and AI background.
Concretely, we created a screening survey with questions about the respondent's background and app usage pattern (e.g., regularly used features, frequency of use). We posted the survey on various channels: Birding International Discord, AI for Conservation Slack, several Slack workspaces within our institution, and Twitter.
On Twitter, in addition to posting the survey, we reached out to accounts with tweets about Merlin via @mentions and Direct Messages.
Based on the screening survey responses, we selectively enrolled participants to maximize the diversity of the study sample's domain and AI background (\cref{tab:participant}).
All participants were active end-users of Merlin who could provide vivid anecdotes of when the app worked well and when it did not.
Regarding the frequency of use, 11 participants used it several times a week, 8 used it once a week, and one used it once a month.

\subsection{Interview sessions and analysis}
\label{sec:interview}

We interviewed 20 participants, over a Zoom video call, from July to August 2022. 
The interviews lasted one hour on average, and we compensated participants with their choice of a 30 USD gift card or a donation to a bird conservation organization made on their behalf. 
As noted in \cref{sec:intro}, in the same interviews, we also inquired about participants' explainable AI needs, uses, and perceptions, and analyzed that portion of the data in another paper~\cite{kim2023helpmehelptheai}. In this work, we focus on understanding what factors influence participants' trust in AI and how. Below we describe the relevant part of the interview protocol. See \cref{sec:app_interviewprotocol} for the interview questions.

We began each interview by introducing the study to the participant, communicating that we were not affiliated with the Merlin development team, and receiving consent for participation in research. We then asked about their domain and AI background, as well as their goals and stakes in their app use.

Next, we inquired about the participant's perception of, experience with, and trust in the app. 
Regarding trust, we adopted Benk et al.'s trust enablement paradigm~\cite{Benk2022TRAIT} and asked participants to describe their trust relationships with the app in their own terms. We asked about general perceptions and attitudes, such as how accurate and trustworthy they find the app, as well as specific instances, such as how they assess the correctness of the app outputs, and in what circumstances they decide to use the app and not. Scoping down our unit of analysis, from the system as a whole to the ``instance'' of use, provided a way to gather dynamic data from our participants about which contextual factors they considered during their trust-related decision-making. 
Finally, we asked the participant whether they would adopt the app in hypothetical high-stakes scenarios with health-related and financial outcomes:
\begin{enumerate}
    \item \textit{Sick bird scenario}: ``Suppose you find a sick bird and take it to the vet. The vet is not sure what bird it is. Would you recommend Merlin to identify the bird species so that the vet can determine the course of treatment?'' We asked the participant to view Merlin as a decision-support tool as the participant and the vet will make the final call.
    \item \textit{Game show scenario}: ``Suppose you enter a game show where you can win or lose money based on how well you can identify birds from photos or audio recordings. You can only use one resource among Merlin, books (e.g., field guides), the Internet (e.g., search engine, online birder community), and so on. Which resource would you use? Does your answer change depending on certain factors?''
\end{enumerate}
We designed these scenarios to introduce high stakes into the AI adoption decision. These scenario-based inquiries allowed us to observe how participants' trust-related assessment and decision-making differ across usage contexts.

We transcribed the interviews and analyzed the transcripts using abductive coding. 
Tavory and Timmermans \cite{tavory2014abductive} describe abductive coding as an iterative process, moving between empirical data and available theory, in order to ensure findings are informed by, in dialogue with, and can contribute to, social-science literature.
We first read five transcripts to identify an initial set of empirical observations through which we could develop a theoretical hunch about the nature of trust-related perceptions, attitudes, and behaviors.
We then consulted the literature on trust and trustworthiness, at which point we found Mayer et al.'s trust model \cite{Mayer1995trust}, which provided a high-level framework and starting point. 
Using this model, we collectively developed a codebook with which we could analyze our initial observations and discern which of our theoretical hunches were novel contributions. 
Finally, we coded all of our data, discussed the results, and drew out themes.

\begin{table}[t!]
    \centering
    \caption{Participants' domain (bird) and AI background. 
    See \cref{sec:app_background} for a description of the categories.}
    \label{tab:participant}
    \begin{tabular}{llll}
    \toprule
    & \textbf{Low-AI} & \textbf{Medium-AI} & \textbf{High-AI} \\
    \midrule
    \textbf{Low-domain} & P7, P12, P16 & P8, P14 & P11, P13 \\
    \textbf{Medium-domain} & P2, P20 & P1, P4, P10 & P6 \\
    \textbf{High-domain} & P5, P17 & P3, P9, P15 & P18, P19 \\
    \bottomrule
    \end{tabular}
\end{table}

\section{Results}
\label{sec:results}

We start by discussing the definition of trust in our study context (\cref{sec:prelude}) and then present multiple aspects of participants' trust in the app: trustworthiness perception and trust attitude (\cref{sec:trustworthiness}); acceptance of individual AI outputs (\cref{sec:acceptance}); and AI adoption decision (\cref{sec:adoption}). Throughout, we note the factors influencing participants' trust, and close with a summary of the results (\cref{sec:summary}).

\subsection{Prelude: Is ``trust'' the right term for participants' attitude toward the app?}
\label{sec:prelude}

Before diving into the findings, let us first examine if ``trust'' is the right term for describing participants' attitudes toward the app. 
Recall from \cref{sec:relatedwork} that trust is defined as an \textit{attitude} and requires \textit{positive expectations} and \textit{vulnerability} in the trustor-trustee relationship \cite{Mayer1995trust,LeeAndSee2004,Vereschak2021CSCW}.
It is easy to see that participants had \textit{positive expectations}: they were actively using the app because they expected it to help them achieve their goal of accurately identifying birds.
However, is there \textit{vulnerability} in use of this everyday app for bird identification?
We answer yes because the app is used in situations involving \textit{uncertainty of outcomes} and \textit{potentially negative consequences}, satisfying \cite{Vereschak2021CSCW}'s definition of \textit{vulnerability}.

First, bird identification is a challenging task that requires the selection of a species among approximately 10,000 existing bird species, some of which are markedly similar to each other.
Even though the app has been developed by bird and AI experts and trained on a large database of expert-annotated bird photos and audio recordings, it is not foolproof.
There is always \textit{uncertainty} about whether it would return an accurate identification, which participants were aware of. See \cref{sec:trustworthiness} for detailed accounts of how participants perceived the app's ability and trustworthiness.

There are also \textit{potential negative consequences} when the app makes a misidentification.
We heard the following responses when we asked participants what they gain and lose when the app succeeds and fails on the task.
As gains, participants mentioned curiosity satisfaction (All), joy \interview{1,7,9,12}, bird knowledge \interview{4,5,8,9,10}, and improved birding experience \interview{1,2,3,4,10}.
As losses, although several participants said \shortquote{nothing material} \interview{1,3,11,12,13,15,16,17}, many expressed that they feel \shortquote{disappointed}, even \shortquote{frustrated}, when the app fails because they really care about correctly identifying birds and would like to gain accurate knowledge about birds \interview{1,4,5,6,9,10,13,15,18,19,20}.
Some noted that misidentifications can lead to people gaining wrong knowledge, (unintendedly) sharing misinformation by reporting wrong bird sightings, and negatively impacting science and conservation efforts \interview{4,5}.

In summary, there were \textit{positive expectations} and \textit{vulnerability} in participants' use of the app, although there were individual differences in the amount of stakes participants placed in their use.
Hence, we conclude ``trust'' is the right term for describing participants' \textit{attitudes} toward the app.
With this established, we now proceed to describe participants' trust in AI in three parts.

\subsection{Trustworthiness perception and trust attitude: Participants assessed the AI to be trustworthy and trusted it}
\label{sec:trustworthiness}

We begin unpacking participants' trust in AI by explaining how they assessed the app's trustworthiness, a key antecedent to trust in Mayer et al.'s model~\cite{Mayer1995trust}.
Overall, participants assessed the app to be trustworthy and trusted it. 
We draw this conclusion based on participants' responses regarding the app's ability, integrity, and benevolence---the three factors of perceived trustworthiness in Mayer et al.'s model (\cref{fig:mayer}). 
Participants assessed that the app possesses all three, based on their prior experience with it, its popularity, and the reputation of the domain and the developers.

\subsubsection{Participants assessed the AI's ability based on their prior experience with the AI and the AI's popularity}

Ability refers to the trustee's skills and competencies \cite{Mayer1995trust}. For automation systems, Lee and See~\cite{LeeAndSee2004} describe it as performance, i.e., how well the automation is performing.
Participants were overall impressed with the app and judged it to have high ability.
Most described the app as very successful and that it seemed to be correct 9-10 out of 10 times, based on their \textbf{prior experience} with the app.\footnote{Recent works~\cite{Banovic2023CSCW,schlicker2022preprint} suggest there may be a gap between the actual and perceived trustworthiness of an AI system. In our study context, however, participants' assessment of the app's ability seems reasonably accurate. Our judgment is based on public knowledge, as we are not affiliated with the app development team. Regarding Photo ID, one of the developers said in 2016 that its ``accuracy is around 90 percent if the user's photo is of good quality''~\cite{perkins_2016}, and we expect the performance would have increased. Regarding Sound ID, recent research from the Cornell Lab of Ornithology suggests that AI models are highly capable of sound-based bird identification~\cite{birdnet}.}
Exceptions were P2, who said Sound ID often made mistakes, and P6, who was disappointed with Photo ID.
Most other participants were very impressed and described the app as \shortquote{pretty insane} \interview{15}, \shortquote{perfect} \interview{11}, and gave high praise despite having observed mistakes: 
\shortquote{I've had one or two times where I've thought I don't believe that's really that bird? [...] But I trust it. I trust it} \interview{14}.
Intriguingly, some participants mentioned they could not accurately assess the app's ability due to their lack of \textbf{domain knowledge} \interview{11,12,13}.
For instance, P11 said: \shortquote{As far as I know, it's been perfect, but I don't know enough to know if it would be making mistakes.}
We discuss this point further in \cref{sec:acceptance}, where we describe how participants assessed the correctness of individual app outputs.
Finally, while most assessed the app's ability based on their own prior experience, 
P12 made an assessment based on the app's \textbf{popularity}, which is an external factor: \shortquote{I imagine that if it has such a wide user base, it would be pretty accurate} \interview{12}.

\subsubsection{Participants assessed the AI's integrity based on the developers' reputation}

Integrity refers to the degree to which the trustee adheres to a set of principles that are acceptable to the trustor \cite{Mayer1995trust}. For automation systems, Lee and See \cite{LeeAndSee2004} describe it as a process, i.e., in what manner and with which algorithms it is accomplishing its objective.
We found that participants believed in the app's integrity because of the \textbf{reputation of the developers}, the Cornell Lab of Ornithology, which is a respected institution with a long history of bird science and conservation efforts.
Most participants were well aware that the app was developed by this lab \interview{1,3,4,5,8,9,11,12,13,14,15,17,18,19}.
Participants were also familiar with the lab's other apps (e.g., eBird, BirdNET, iNaturalist) and resources (e.g., All About Birds, Macaulay Library), describing these and the app as their \shortquote{go-to} \interview{15} when they want to learn about a specific bird.
P14 specifically said they trusted the app because it was developed by this lab: \shortquote{I know that Cornell Ornithology Lab does excellent, excellent stuff. I mean, if you're going to try and learn anything about a bird, just go there. Don't try anything else. Don't even bother the Audubon Society. Just go straight to Cornell. So I trusted it [the app] for that reason.}
Participants did not know how the app was developed or how it works, since such information is not publicly available. Nonetheless, they believed in the app's integrity because they believed in the authority and expertise of the app developers.

\subsubsection{Participants assessed the AI's benevolence based on the domain's reputation}

Benevolence refers to the extent to which the trustee’s motivations are aligned with the trustors' \cite{Mayer1995trust}. For automation systems, Lee and See \cite{LeeAndSee2004} describe it as purpose, i.e., why the automation was built originally. 
We found that participants believed in the app's benevolence because of the positive \textbf{reputation of the domain}, i.e., the birding community that they and the app developers are part of.
For instance, P18 described the birding community as a place where everyone tries to be accurate and do good: \shortquote{I think birders, in general, are a community where there's very few people who try and do adversarial attacks because it doesn't benefit anybody [...] the value of the birding community is that everybody is trying to be accurate.}
Some participants contrasted the app with other AI applications.
For instance, P2 described the app as not having \shortquote{malicious intent} compared to advertisements.
P9 contrasted the app with other AI applications they found \shortquote{creepy} and \shortquote{harmful,} such as voice assistants that may be \shortquote{monitoring} user behavior.
Although the app may also collect personal location data, participants seemed less concerned overall.
P18 and P19 even wanted their data to go somewhere and be used (e.g., for science and conservation efforts or improving the app) so that it is \shortquote{contributing something to society} \interview{19}.

\subsection{Output acceptance: Participants selectively accepted AI outputs after verification}
\label{sec:acceptance}

Participants described the app as trustworthy and trusted it; however, they did not accept its outputs as true in every single instance of use.
To the extent possible, participants carefully assessed the app's outputs, using their knowledge about the domain, and then made acceptance decisions.
Our findings reveal a gap between \textit{general} trustworthiness perceptions and trust attitudes, and \textit{instance-specific} trust-related behaviors, highlighting the importance of considering both aspects in trust in AI research.

\subsubsection{Participants assessed AI outputs based on their likelihood and task difficulty}

Participants had developed heuristics for judging the correctness of app outputs \interview{1,2,6,16,17,19}.
One heuristic was assessing the \textbf{likelihood} of spotting a bird species in a given area.
Participants described they were more trusting of the app's output when the identified species is common for the area, and less trusting when it is rare.
For instance, P6 said they judge the output's correctness based on recent sightings in the area and rarity: \shortquote{If it's a common bird or even just a rare bird, uncommon or something like that, then maybe [it is correct]. But if it's a super rare bird, then definitely not.}
Another heuristic concerned \textbf{task difficulty}.
P1's response well explains this heuristic: \shortquote{I trust it [the app] more when I know that I'm looking at something that should be relatively unambiguous. If I'm looking at something that's like a Female Warbler or a Female Sparrow, which might just be a little brown bird, then I'm a little bit more skeptical of the result.}
For context, \shortquote{little brown bird} is a term used by birders to describe a large number of species of small brown passerine birds, which are known to be notoriously difficult to distinguish. P1 described them as \shortquote{really hard to ID, even for a human ornithologist.}
We note that both heuristics require \textbf{domain knowledge} as participants must know whether the identified species is common or rare for the area, and which birds are difficult and easy to identify.

\subsubsection{Participants verified AI outputs with input-output comparisons and information from other sources}

Some participants verified the outputs via \textbf{input-output comparisons}, i.e., they compared input photos and audio recordings to reference photos and audio recordings of the identified species, which are provided in the app \interview{1,10,20}.
For instance, P10 verified Photo ID outputs as follows: \shortquote{I go back and look at reference photos of that bird and then try to map field markings. So I'll try to see if the bill shape lines up.}
P20 described a similar process for verifying Sound ID outputs: \shortquote{They have that drop down [menu] that has the different sounds, and I will literally just play one until I find one that I'm like: oh it's that, that's the exact sound.}
In short, participants looked for the resemblance between the inputs and the references, and then assessed the outputs' correctness.
This verification does not require \textbf{domain knowledge} per se; however, participants with it could more easily verify the output as they would know what to check.

Participants also described using \textbf{information from other sources} \interview{1,4,10,15,18}.
If the app identified a bird based on sound, participants tried to confirm it with their own visual identification, and vice versa \interview{10,18}.
For instance, P10 said they try to visually confirm the bird when using Sound ID: \shortquote{I'll also look for that bird and see if I can see it as well. Or see if it matches a bird that I have seen.}
Some participants took a step further and consulted other birders, through their personal networks or online communities \interview{1,4,15}.
P1 said they often share the app's outputs with friends and birders online. They also remarked, \shortquote{If there was something that I knew was more of an ambiguous result, I would usually consult somebody} \interview{1}.
P4 was an especially active member of online communities. They said they ask questions on Discord and Reddit and consult expert birders in the area: \shortquote{I do oftentimes reach out to people whose names I find on e-bird. I see local checklists and I'll either find them on Facebook or LinkedIn or something and send them a message.}
Note that cross-checking requires \textbf{domain knowledge} for identifying birds on their own, whereas consulting other birders does not.

\subsubsection{Some participants disregarded AI outputs when they could not verify}

For some participants, verification was a crucial and necessary step for output acceptance \interview{3,4,15,18}.
When unable to verify, they disregarded the app output.
For example, P15 said they've never only relied on the app when identifying a bird they have not seen before. They almost always sent the output to more experienced birders and received their confirmation.
P18 was also strict about when they accept app outputs, stating, \shortquote{I never, I never count on my bird registry anything that Sound ID says that I can't kind of confirm either through the facts of it or through a visual ID} \interview{18}.
These participants disregarded unverifiable app outputs, despite their positive assessment of the app's ability and trustworthiness, revealing a gap between general trustworthiness perceptions and instance-specific trust-related behaviors.

\subsubsection{Not all participants had the ability to assess the correctness of AI outputs}

So far we described various processes through which participants decided whether or not to accept app outputs.
However, not all participants had the \textbf{ability to assess} the correctness of app outputs.
In \cref{sec:trustworthiness}, we described how some participants with little \textbf{domain knowledge} said they could not accurately assess the app's ability \interview{11,12,13}.
These participants also said that because they \shortquote{know so little about birds} \interview{12}, they could not \shortquote{validate or reject} \interview{11} app outputs, especially if they can't get information from other sources.
P13 said, \shortquote{If it's misidentifying a bird that I can't see, then I have no way to know that.}
This finding suggests that domain knowledge is a key factor in appropriate trust calibration and has a wide influence on participants' interactions with the app.

\subsection{Adoption: Participants never decided against AI adoption in their actual use setting, but made different decisions for hypothetical high-stakes scenarios}
\label{sec:adoption}

The final aspect of trust in AI we analyze is how participants made AI adoption decisions.
We compared participants' decision-making process
between their actual use setting and two hypothetical high-stakes scenarios (see~\cref{sec:interview} for the scenario details).
We found that while participants always used the app in their actual use setting, they made different adoption decisions for the high-stakes scenarios based on various factors: the app's ability, familiarity, and ease of use (AI-related factors); participants' ability to assess the app's outputs and use the app (Human-related factors); and finally, task difficulty, perceived risks and benefits of the situation, and other situational characteristics (Context-related factors).

\subsubsection{In their actual use setting, participants never decided against using the AI}

We found that participants always use the app when opportunities arise.
It is not that participants absent-mindedly used the app.
Participants were aware of when the app works well and not, and knew how to help the app be more successful, e.g., by supplying better input photos and audio recordings.
However, when we asked how they make app adoption decisions, they only described situations where they decided to use the app, and never situations where they decided against using it.

There could be several reasons for this finding.
First, the app has a low cost of use. Since the app is free, the only use costs are the time and effort involved in taking photos or audio recordings and inputting them into the app, and perhaps a small amount of phone battery.
Second, the risks of use are also low. There are potential negative consequences when the app misidentifies, e.g., gaining wrong knowledge, as described in \cref{sec:prelude}.
However, end-users can mitigate these risks by verifying the output and rejecting it if needed.
Finally, we only interviewed active end-users of the app, who are likely to continue to use the app because they are satisfied with it. Past or non-users may provide different responses.

\subsubsection{In hypothetical high-risk scenarios, participants carefully considered the AI's ability and various contextual factors}

When we presented participants with hypothetical high-risk scenarios, we observed a different decision-making process around app adoption.
Participants considered the app's \textbf{ability} with respect to various \textbf{situational characteristics}.
For example, for the sick bird scenario, some participants judged the app is worth a try.
P15 described using the app as \shortquote{something that wouldn’t hurt} since they are in a situation where both they and the vet could not identify the bird. They expressed some degree of confidence in the app's ability: \shortquote{I feel like Merlin's not gonna tell you that a baby hawk is a chickadee} \interview{15}.
Similarly, P17 said they \shortquote{would definitely recommend it [the app] to get into the right ballpark.} Still, they recommended consulting other scientific resources and doing a \shortquote{triple check} of the app's output, since the \textbf{risk} of misidentification, i.e., the sick bird receiving the wrong treatment, is higher than the risk in their actual use settings, e.g., gaining wrong knowledge.

Other participants were skeptical that the app could identify the sick bird \interview{3,6,15,19}.
P4 did not think the app could identify birds that they and the vet could not:
\shortquote{Assuming that I don't know what the bird is and they [vet] don't know what the bird is, this bird is some ambiguous-looking bird. In those cases [...] I don't think Merlin would be able to know.}
P15 pointed out that sick birds are often \shortquote{fledglings, juveniles} which are \shortquote{harder to ID for everybody in real life and presumably harder for Merlin.}
P6 noted that sick birds may be ``out-of-distribution'' for the app due to their underrepresentation in the training data: \shortquote{I assume Merlin is not trained on sick birds, so I can totally see it doing something crazy.}
These participants weighted the app's ability against the \textbf{task difficulty} and decided against adopting the app in the sick bird scenario.

Similarly, for the game show scenario, participants jointly considered the app's ability and the situation's characteristics.
P1 and P4 said they would choose the app if there are \textbf{time constraints}, but otherwise choose \shortquote{a good quality field guide} \interview{1} or \shortquote{Discord} \interview{4}.
Similarly, P15 said they would choose the app if they need to give an answer \shortquote{quickly, within 30 seconds} but otherwise consult other birders.
Others said they would choose the app if they have to do \textbf{sound-based identification} on the game show \interview{6,10,12,20}, describing difficulties with text-based referencing of sound: \shortquote{some books saying `it goes da-da-da' is not helpful} \interview{6}.
P20 explained their reasoning in detail: \shortquote{If I'm on this game show and it plays a sound, I would definitely want to use the app [...] but if it shows me a bird, I might just want to google it because I have enough knowledge personally that I could probably guess what type [...] and then search by colors. So I guess it comes down to what I think the app does the best, which is sound, versus what I think I can get away without it.}
Participants considered the app's ability not only on its own, but also in comparison to other resources.
Again, participants carefully made app adoption decisions as the \textbf{perceived risks and benefits} of the scenario, i.e., loss and gain of money, are higher than those of their actual use setting.

\begin{table*}[t!]
    \centering
    \caption{Factors that influenced our study participants' trust in AI. See \cref{sec:summary} for a discussion.}
    \label{tab:factors}
    \begin{tabular}{lll}
    \toprule
    \textbf{Human-related factors} & \textbf{AI-related factors} & \textbf{Context-related factors} \\
    \midrule
    Domain knowledge & Ability & Task difficulty \\
    Ability to assess the AI's outputs & Integrity & Perceived risks and benefits \\
    Ability to assess the AI's ability & Benevolence & Situational characteristics \\
    Ability to use the AI & Popularity & Domain's reputation \\
     & Familiarity & Developers' reputation \\
     & Ease of use & \\
    \bottomrule
    \end{tabular}
\end{table*}

\subsubsection{Some participants adopted the AI due to familiarity and ease of use}

Two other factors that impacted participants' app adoption decisions were \textbf{familiarity} and \textbf{ease of use}.
For the sick bird scenario, P2 said they would definitely used the app because it \shortquote{feels kind of like second nature.}
P16 also chose to identify the sick bird with the app because using the app \shortquote{would be the easiest.}
Similarly, for the game show scenario, P2 picked the app as their top choice because \shortquote{it's so easy [...] it doesn't take all that much time to look through everything.}
P16 mentioned both familiarity and ease of use: \shortquote{I think Merlin would make the most sense since I'm familiar with it.}
They described other resources as requiring more ``work'' by end-users, compared to the app where end-users can just input bird photos and/or audio recordings: \shortquote{You still have to do a lot of work to do like a Google search compared to this [app]} \interview{16}.

P4 described another aspect of familiarity: their \textbf{ability to use the app}.
They said, \shortquote{I definitely would use Merlin because I'm familiar with it. And I trust my ability, like I know how to operate it pretty well} \interview{4}.
We found this response particularly interesting because the ability to use the AI has not been explored much in the trust in AI literature. 
However, we expect it will become an important topic in trust and human-AI interaction research, as AI applications grow in complexity and require end-users to develop skills for effective use of AI.

Finally, some participants with little \textbf{domain knowledge} said they would adopt the app because other resources lack familiarity and ease of use \interview{11,17}.
For instance, P17 described field guides as a more advanced and less accessible resource than the app \shortquote{because of the way they're structured and organized.}
They said the app will get them closer to an answer \shortquote{a lot quicker} \interview{17}.
Similarly, P11 mentioned they could not effectively use the Internet search engine because they do not know enough about birds to effectively describe the bird they have to identify. 
They added, \shortquote{Maybe someday, when I know a lot about birds, I would feel comfortable using another resource that's more of an expert than Merlin is, but I think Merlin is the right level of expertise for what I know right now} \interview{11}.

\subsection{Summary of the results}
\label{sec:summary}

In short, we found that end-users' trust relationship with AI is complex.
Overall, participants found the app trustworthy and trusted it.
Still, they carefully assessed the correctness of individual outputs and decided against app adoption in certain high-stakes scenarios.
This finding illustrates that trust is a multifaceted construct that must be approached holistically. 
To get a full and accurate picture of trust, it is crucial to examine both \textit{general} aspects such as trustworthiness perceptions and trust attitudes and \textit{instance-specific} aspects such as AI output acceptance and adoption decisions.

We also highlight that trust in AI is influenced by many factors.
In \cref{tab:factors}, we organize the factors we identified based on whether they are related to the human trustor, the AI trustee, or the context, following prior work~\cite{Hartmann2022AMCIS,Jermutus2022healthcare,Kaplan2021review}.
\textbf{Human-related factors} include domain knowledge and other factors influenced by it, such as the ability to assess the AI's outputs, the ability to assess the AI's ability, and the ability to use the AI.
\textbf{AI-related factors} include internal factors such as ability, integrity, and benevolence; external factors such as popularity; and user-dependent factors such as familiarity and ease of use.
\textbf{Context-related factors} include task difficulty, perceived risks and benefits of the situation, other situational characteristics, and the reputation of the domain and the developers.
We emphasize that this is not a complete set of factors that can influence trust in AI, but what we \textit{observed} in our case study in a bottom-up manner.

\section{Discussion}
\label{sec:discussion}

In this section, we discuss the implications of our findings (\cref{sec:discussion_implications}), reflect on the applicability of Mayer et al.' trust model~\cite{Mayer1995trust} (\cref{sec:discussion_mayer}), examine the limitations of our work and opportunities for future work (\cref{sec:discussion_limitations}), and provide practical recommendations for future research on trust in AI (\cref{sec:discussion_recommendations}).

\subsection{Key findings and their implications}
\label{sec:discussion_implications}

\subsubsection{Insights from instance-specific trust-related behaviors}

Participants' instance-specific decisions about AI output acceptance and adoption were particularly useful for understanding what factors influence trust in AI and how.
In \cref{sec:acceptance}, we described how participants trusted the app output more when the task is easy (e.g., \shortquote{relatively unambiguous} bird) and less when the task is difficult (e.g., \shortquote{little brown bird}).
Similarly in \cref{sec:adoption}, we described how some participants were hesitant to use the app to identify the sick bird in the first hypothetical scenario because they judged the task would be too difficult for the app.
These examples illustrate the rich reasoning behind participants' trust-related behaviors, where factors of trust interact with each other.
Participants used their domain knowledge (human-related factor) to assess task difficulty (context-related factor) and weighted it against the app's ability (AI-related factor) to decide whether or not to accept the app output or adopt the app in the given situation.

In earlier sections, we also described domain knowledge's influence on participants' ability to assess the app's ability (\cref{sec:trustworthiness}) and outputs (\cref{sec:acceptance}).
Participants with domain knowledge assessed the correctness of an app output by, for example, cross-checking it with their own identification and judging its likelihood based on their knowledge of what birds are common and rare in the area and what birds are easy and difficult to identify.
Participants without domain knowledge, however, had difficulties in assessing the correctness of app outputs and, consequently, the app's overall ability.

Taken together, these findings have two important implications.
First, they imply that domain knowledge can be a widely influential factor of trust.
In the above examples, participants assessed task difficulty based on their domain knowledge of which birds are difficult and easy to identify.
Further, participants' domain knowledge directly influenced their ability to verify the app's outputs and overall ability, which can have far-reaching influences on participants' trust and interaction with the app.
Second, they imply that participants expected the app to struggle on similar tasks as humans do, when current research suggests AI models do visual recognition differently from the human brain~\cite{Ullman2016PNAS} and make different mistakes than humans~\cite{tuli2021cogsci,xu2019adversarial,koh2021wilds}. However, we cannot rule out the possibility that the app struggles on similar bird identification tasks as humans do, since we do not have access to its underlying AI models. This implication calls for further research on the relationship between end-users' domain knowledge, perceptions and expectations toward AI, and trust calibration ability.

\subsubsection{The surprising impact of domain knowledge}

As aforementioned, participants' domain knowledge had a wide impact, influencing all aspects of their trust in the app.
We found this surprising because domain knowledge has not been discussed much as a factor of trust in the trust in AI literature.
Upon reflection, we speculate this is because most AI applications are either (1) non-expert applications that do not have a notion of domain expertise (e.g., image search) or (2) expert applications that are developed to be used by domain experts only (e.g., clinical decision-support tool).
Our study application is unique in that it is an expert application that is used by both domain experts and non-experts.
Further, our study design, specifically our choice to recruit participants with varying background (\cref{tab:participant}), allowed us to observe group differences with respect to domain knowledge.

This observation has particularly important implications for expert AI applications that support high-stakes decisions about people. Using clinical decision-support tools as an example, we expect domain knowledge differences to lead to very different trust-related assessments and decision-making between experts (clinicians) and non-experts (patients, regulators, and other stakeholders). As we saw in our case study, non-experts may not be able to spot the AI's mistakes and assess its trustworthiness as accurately as domain experts. Even though non-experts are not the intended end-users of this AI (clinical decision-support tool), it is important that they have mechanisms to appropriately calibrate their trust in the AI. Hence, we urge the community to consider domain knowledge when designing AI applications and trust calibration interventions. For example, a system could assess the verification behaviors used by domain experts, and build these options into the system so that they are accessible to experts and non-experts alike.

\subsubsection{The importance of contextual factors and contextually-grounded studies}

Finally, we highlight the importance of contextual factors and contextually-grounded studies for understanding their influences on trust.
When participants were describing the app's trustworthiness, we observed that the positive reputation of the domain (birding community) and the developers (Cornell Lab of Ornithology) led them to positively assess the AI's ability, integrity, and benevolence (\cref{sec:trustworthiness}).
We have two points of discussion on this finding.
First, it shows that external contextual factors (reputation of the domain and the developers) influence internal AI factors (ability, integrity, and benevolence), underlining the impact of contextual factors on trust.
It also reiterates that factors influence each other, and calls for research that studies the interactions between factors.
Second, while this specific finding is context-dependent, it provides generalizable insights.
For example, we can anticipate end-users to have doubts about an AI application's ability if the AI is not developed by a well-known institution; benevolence if the AI seems to have a different goal from them (e.g., recommendation systems trying to sell unneeded products); and integrity if the AI seems to make decisions with wrong reasons (e.g., decision-making systems discriminating based on protected attributes).

\subsection{Adapting existing trust models to AI}
\label{sec:discussion_mayer}

In this work, we used Mayer et al.'s model for organizational trust~\cite{Mayer1995trust} to analyze the empirical data we collected on trust in AI. Overall, we found the model applicable and useful for understanding trust in AI.
In particular, we found helpful the way in which it breaks down ``trust'' into multiple components and delineates trust from its antecedents, context, and products (\cref{fig:mayer}). However, as with any model, there were some limitations and challenges in its application. First, since the model was originally developed for \textit{trust between people}, we had to make adaptations to apply it to \textit{trust in AI}. For instance, when describing participants' trustworthiness perceptions (\cref{sec:trustworthiness}), instead of using Mayer et al.'s \cite{Mayer1995trust} definitions of ability, integrity, and benevolence, we used Lee and See's~\cite{LeeAndSee2004} automation-friendly translations of these factors: performance, process, and purpose. Second, Mayer et al.'s model \cite{Mayer1995trust} is by no means a comprehensive trust model. This is expected as the work's goal was not to list all possible antecedents of trust. 
Hence, we drew from other works~\cite{Kaplan2021review,Hartmann2022AMCIS,Jermutus2022healthcare,Hoff2015model} to categorize the factors we identified into human, AI, and context-related factors. Third, in our study, we did not observe the influence of trustor's propensity, one of the model components. However, this result does not imply that trustor's propensity is an unimportant factor of trust in AI. Future research, in particular survey and experimental studies, are needed for such conclusions.

\subsection{Limitations and future work}
\label{sec:discussion_limitations}

Our work has the following limitations.
First, as with any case study, our findings are context-specific. However, the gained insights, e.g., the trust-influencing factors we identified in \cref{tab:factors}, may generalize to other settings.
Further, our holistic approach to trust may aid future research on other types of AI applications.
Another limitation is that all participants were active end-users of the app. Those who just started using it, stopped using it, or chose to not use it are not represented in the study.
We encourage a more comparative study design for future work so that findings can be compared across non-users and user subgroups.
Finally, due to the highly multifaceted and dynamic nature of trust in AI, there are aspects of it that our work does not cover. 
More research needs to be done, especially on how trust is initially developed and changes overtime, and how trust relationships with AI vary between stakeholder groups.

We highlight more important areas for future work.
In our view, the overarching goal of trust in AI research is to establish \textit{warranted and calibrated trust in AI}, where people's trust in AI systems matches their actual trustworthiness.
We see three steps of research to achieving this goal.
The first is to deepen the understanding of trust in AI, e.g., what aspects there are to trust and what factors influence it.
Our work falls into this first step.
The second is to understand how different factors influence trust, likely with quantitative methods such as surveys and behavioral experiments.
We urge the field to move from studying one or a few factors in lab settings with hypothetical end-users, to studying multiple factors in real-world settings with actual end-users. This shift is necessary for understanding the interactions between factors, as well as the contextual influences on trust.
The third step is to design effective trust calibration interventions, based on the gained insights on trust and its influencing factors.
We point to Wischnewski et al.'s work \cite{wischnewski2023chi} for a survey of the state-of-the-art trust calibration interventions and suggestions for future directions.

\subsection{Practical recommendations}
\label{sec:discussion_recommendations}

We conclude with a set of practical recommendations for future research on trust in AI.
\begin{enumerate}
    \item \textit{State a definition of trust.} Trust is a multifaceted construct that carries different meanings to different people. Explicitly stating a definition of trust can help remove confusion around the term and encourage accurate interpretation and comparison of study results.
    \item \textit{Examine if trust is the construct being studied.} 
    Oftentimes what's being studied is not trust, but other related constructs such as confidence and reliance, as we discussed in \cref{sec:relatedwork}.
    We recommend that researchers carefully examine their study design and context to ensure trust is the construct being analyzed. We hope our \cref{sec:prelude} serves as a helpful example of such an examination.
    \item \textit{Approach trust holistically and study its antecedents, context, and products.} Our contextually-grounded study of general trustworthiness perceptions and trust attitudes, as well as instance-specific trust-related behaviors, revealed a comprehensive picture of end-users' trust relationships with AI that cannot be gained by studying only one aspect of trust. Hence, we recommend studying trust together with its antecedents, context, and products, to the extent possible.
    \item \textit{Consider human, AI, and context-related factors and their interactions.} As observed in this work, trust is influenced by many factors. To prevent surprises and gain a thorough understanding of trust in a given context, we recommend anticipating as many factors as possible and studying their interactions. We found it particularly helpful to consider factors along the dimensions of human, AI, and context.
\end{enumerate}

\section{Conclusion}
\label{sec:conclusion}

We conducted a qualitative descriptive empirical study of end-users' trust in AI in a real-world context. We interviewed 20 end-users of a widely-used AI-based app for bird identification~\cite{merlin} and inquired about their trust in the app from many angles. Using a process-oriented trust model~\cite{Mayer1995trust}, we elaborated on multiple aspects of trust in AI. Notably, we found a discrepancy between participants' \textit{general} trustworthiness perceptions and trust attitudes, and \textit{instance-specific} trust-related behaviors, adding nuances to existing understandings of trust in AI. We also identified human, AI, and context-related factors of trust, finding that domain knowledge had a particularly big influence on participants' trust and interaction with AI. Finally, we discussed the implications of our findings and provided recommendations for future research toward establishing warranted and calibrated trust in AI. We hope our work aids future research on other AI applications and the various contexts into which they are integrated.

\begin{acks}
We foremost thank our participants for generously sharing their time and experiences.
We also thank Tristen Godfrey, Dyanne Ahn, and Klea Tryfoni for their help in the interview transcription.
Finally, we thank the anonymous reviewers and members of the Princeton HCI Lab and the Princeton Visual AI Lab (especially Angelina Wang, Vikram V. Ramaswamy, Amna Liaqat, and Fannie Liu) for their helpful and thoughtful feedback.
This material is based upon work partially supported by the National Science Foundation (NSF) under Grants No. 1763642 and 2145198 awarded to OR. Any opinions, findings, and conclusions or recommendations expressed in this material are those of the authors and do not necessarily reflect the views of the NSF.
We also acknowledge support from the Princeton SEAS Howard B. Wentz, Jr. Junior Faculty Award (OR), Princeton SEAS Project X Fund (RF, OR), Princeton Center for Information Technology Policy (EW), Open Philanthropy (RF, OR), and NSF Graduate Research Fellowship (SK).
\end{acks}

\bibliographystyle{ACM-Reference-Format}
\bibliography{references}

\clearpage

\appendix
\section*{Appendix}

\section{Domain and AI background levels}
\label{sec:app_background}

We provide details on the background levels mentioned in \cref{sec:recruitment} and \cref{tab:participant} of the main text.
We grouped participants based on their survey responses and interview answers.
\begin{itemize}
    \item \textit{Low-domain}: From ``don't know anything about birds'' (P11, P12) to ``recently started birding'' (P7, P8, P13, P14, P16). Participants who selected the latter option typically have been birding for a few months or more than a year but in an on-and-off way, and were able to identify some local birds.
    \item \textit{Medium-domain}: Have been birding for a few years and/or can identify most local birds (P1, P2, P4, P6, P10, P20).
    \item \textit{High-domain}: Have been birding for more than a few years and/or do bird-related work (e.g., ornithologist) (P3, P5, P9, P15, P17, P18, P19).
    \item \textit{Low-AI}: From ``don't know anything about AI'' (P16, P17) to ``have heard about a few AI concepts or applications'' (P2, P5, P7, P12, P20). Participants in this group either did not know that the app uses AI \interview{12,16} or knew but weren't familiar with the technical aspects of AI \interview{2,5,7,17,20}.
    \item \textit{Medium-AI}: From ``know the basics of AI and can hold a short conversation about it'' (P1, P3, P8, P9, P14) to ``have taken a course in AI or have experience working with an AI system'' (P4, P10, P15). Participants in this group had a general idea of how the app's AI might work, e.g., it is neural network based and has learned to identify birds based on large amounts of labeled examples.
    \item \textit{High-AI}: Use, study, or work with AI in day-to-day life (P6, P11, P13, P18, P19). Participants in this group were extremely familiar with AI in general and had detailed ideas of how the app's AI might work at the level of specific data processing techniques, model architectures, and training algorithms.
\end{itemize}
Note that our referral here and elsewhere to ``high-AI background'' participants describes their expertise with AI in general, not necessarily with the app's AI.

\section{Interview Protocol}
\label{sec:app_interviewprotocol}

As we noted in the main text, we used interview data from our prior study about Explainable AI~\cite{kim2023helpmehelptheai}.
Below is the portion of the interview protocol we used to understand what factors influence participants' trust in AI and how.

\subsubsection*{Domain and AI background}
\begin{enumerate}
    \item How would you describe your knowledge of birds?
    \item How would you describe your knowledge of machine learning and artificial intelligence?
\end{enumerate}

\subsubsection*{Use of app}
\begin{enumerate}
    \item Which features of Merlin do you use among Bird ID, Photo ID, Sound ID, Explore Birds? Why do you not use features XYZ?
    \item For what tasks do you use the app?
    \item How successful are you at accomplishing those tasks?
    \item In what scenarios or circumstances do you decide to use the app?
\end{enumerate}

\subsubsection*{Stakes in use}
\begin{enumerate}
    \item What do you gain when Merlin is successful? What do you lose when Merlin is unsuccessful?
    \item How important is it to you that Merlin gets each and every prediction correct?
\end{enumerate}

As you may know, Merlin uses machine learning-based AI models to identify birds in photos and audio recordings. We will now ask questions about your experiences and thoughts on Merlin’s AI models.

\subsubsection*{Knowledge and perception of AI}
\begin{enumerate}
    \item What do you know about Merlin’s AI?
    \item How accurate do you think Merlin's bird identification is? 
    \item How well did you expect Merlin to work? How well did it actually work?
    \item How do you know if Merlin is correct or incorrect?
    \item Do you know when Merlin works well and not? For example, have you noticed that it works better on certain types of inputs or certain bird species?
\end{enumerate}

\subsubsection*{High-risk scenarios}

We will now present two scenarios to you and ask whether you would use Merlin in them.
\begin{enumerate}
    \item Scenario 1: Suppose you find a sick bird and take it to the vet. The vet is not sure what bird it is. Would you recommend Merlin to identify the bird species so that the vet can determine the course of treatment?
    \item Scenario 2: Suppose you are entering a game show where you can win or lose money based on how well you can identify birds from photos or audio recordings. You can only use one resource among Merlin, books (e.g., field guides), the Internet (e.g., search engine, online birder community), and so on. Which resource would you use? Does your answer change depending on certain factors?
\end{enumerate}

\subsubsection*{Closing}
Is there anything that you want the research team to know that we haven’t been able to cover yet?

\end{document}